\documentclass[aps,prl,twocolumn,superscriptaddress,showpacs]{revtex4-1}

\usepackage{graphicx}
\usepackage{amsmath,amssymb}

\begin{document}

\title{Low Temperature Dynamics of Magnons in a Spin-1/2 Ladder Compound}
\author{B.~N\'afr\'adi}
\affiliation{Max-Planck-Institut f\"ur Festk\"orperforschung, Heisenbergstra\ss e 1, D-70569 Stuttgart, Germany}
\author{T.~Keller}
\affiliation{Max-Planck-Institut f\"ur Festk\"orperforschung, Heisenbergstra\ss e 1, D-70569 Stuttgart, Germany}
\affiliation{ZWE FRM II, TU M\"unchen, Germany}
\author{H.~Manaka}
\affiliation{Graduate School of Science and Engineering, Kagoshima University, Kagoshima 890-0065,Japan}
\author{A.~Zheludev}
\affiliation{Neutron Scattering and Magnetism Group, Laboratorium f\"ur Festk\"orperphysik, ETH Z\"urich, CH-8093, Switzerland}
\author{B.~Keimer}
\affiliation{Max-Planck-Institut f\"ur Festk\"orperforschung, Heisenbergstra\ss e 1, D-70569 Stuttgart, Germany}

\date{\today}

\begin{abstract}
We have used a combination of neutron resonant spin-echo and triple-axis spectroscopies to determine the energy, fine structure, and linewidth of the magnon resonance in the model spin-1/2 ladder antiferromagnet IPA-CuCl$_3$ at temperatures $T \ll \Delta_0 /k_B$, where $\Delta_0$ is the spin gap at $T=0$.
In this low-temperature regime we find that the results deviate substantially from the predictions of the non-linear sigma model proposed as a description of magnon excitations in one-dimensional quantum magnets and attribute these deviations to real-space and spin-space anisotropies in the spin Hamiltonian as well as scattering of magnon excitations from a dilute density of impurities.
These effects are generic to experimental realizations of one-dimensional quantum magnets.

\end{abstract}


\maketitle

The concept of weakly interacting ``quasiparticles'' underlies much of our understanding of condensed matter physics.
One-dimensional (1D) spin liquids, such as spin-1 chains or spin-1/2 ladders, have emerged as important model systems for this concept, because they are amenable to a comprehensive theoretical description.
These systems exhibit collective singlet ground states, and the quasiparticles are spin-1 triply degenerate ``magnons'' with a gap $\Delta_0$ at temperature $T \rightarrow 0$.
Thermally excited magnons cannot avoid collisions in 1D, but they remain well-defined quasiparticles as long as their density is low and collisions are rare, which is the case for $k_B T \ll \Delta_0$.
In this low-temperature regime, theoretical work based on the non-linear sigma model (NL$\sigma$M) has yielded a universal description of the quasiparticle properties \cite{Affleck1989,Jolicoeur1994,Sachdev1997,Damle1998}.
The theory predicts that the temperature dependent energies and lifetimes of magnons depend solely on $\Delta_0$, and it has yielded analytical expressions for these observables.
The complete theoretical description of quasiparticles obtained in this way is unique in condensed matter physics.

Unfortunately, experimental access to this universal low-temperature regime has been severely limited.
Thermodynamic, thermal transport \cite{Kordonis2006,Hess2007,Sologubenko2008}, and nuclear magnetic resonance (NMR) \cite{Takigawa1996,Ghoshray2005} experiments have provided valuable insights into the relaxation of magnons in materials containing quasi-1D networks of quantum spins.
However, in gapped spin liquids these methods provide only limited information at low temperatures, because their sensitivity scales with the magnon density that vanishes exponentially as $T \rightarrow 0$.
NMR experiments have revealed substantial deviations from the predictions of the NL$\sigma$M at the lowest temperature accessible to this method \cite{Takigawa1996,Ghoshray2005}, but they do not provide the spectral resolution required to uncover whether these deviations reflect a fundamental inadequacy of the NL$\sigma$M, or whether additional factors such as spin-space anisotropies, inter-chain or inter-ladder interactions, or scattering from impurities are responsible.
Electron spin resonance (ESR) provides spectral resolution in the $\mu$eV range \cite{Manaka2000}, but like NMR this technique relies on thermal excitation of magnons and is therefore limited to $k_B T \gtrsim \Delta_0$.
Inelastic neutron scattering (INS) creates magnons in the scattering process and is therefore applicable at all temperatures, at least in principle.
INS experiments performed on a variety of quasi-1D magnets with gapped excitation spectra have indeed found good agreement of the measured magnon lifetimes with the predictions of the NL$\sigma$M over a surprisingly large temperature range \cite{Xu2007,Zheludev2008}.
Due to insufficient energy resolution, however, it has thus far not proven possible to extend these experiments into the low-temperature regime where the NL$\sigma$M approach is rigorously justified.

Recent experiments have shown that a combination of neutron resonant spin-echo (NRSE) and triple-axis (TAS) spectroscopies can enhance the energy resolution of conventional INS by about two orders of magnitude, by taking advantage of the Larmor precession of neutron spins in radio-frequency magnetic fields \cite{Bayrakci2006,Keller2006,Aynajian2008}.
In NRSE-TAS experiments, neutrons create gapped magnons in the same manner as conventional INS.
This method thus combines the strengths of ESR and INS in such a way that experiments with spectral resolution in the $\mu$eV range can be performed at arbitrarily low temperatures.

\begin{figure}
	\includegraphics[width=7.9cm]{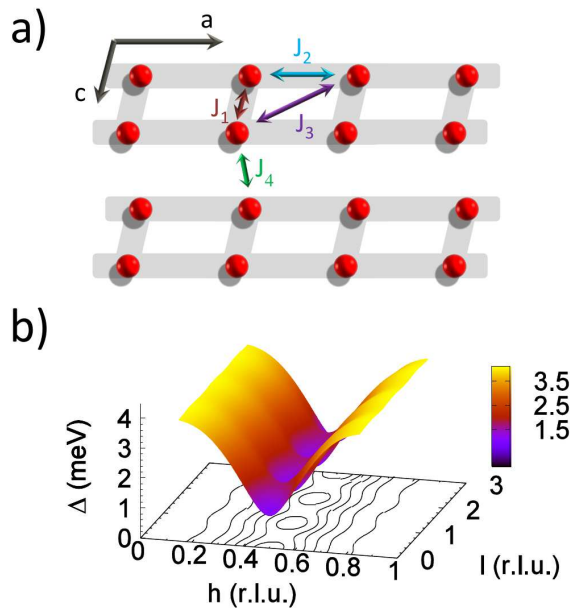}
   \caption{(Color online) a) Schematic view of the \textit{a-c} plane of IPA-CuCl$_3$. The red balls represent spin-1/2 Cu$^{2+}$ ions, the arrows show the dominant exchange interactions as discussed in the text. b)  Color plot of the magnon dispersion from Ref. \onlinecite{Masuda2006} \label{fig:IPA}}
\end{figure}

We report NRSE-TAS experiments on the model spin-1/2 ladder system IPA-CuCl$_3$ (where IPA denotes isopropyl ammonium, (CH$_3$)$_2$CHNH$_3$), one of the most extensively studied quasi-1D quantum spin liquid \cite{Manaka2000,Masuda2006,Zheludev2007,Garlea2007,Zheludev2008}.
In contrast to prior INS work, the spectral resolution is sufficient to resolve a subtle splitting of the magnon resonance by anisotropic exchange interactions.
We find that the temperature dependencies of the magnon lifetime and energy deviate strongly from the predictions of the NL$\sigma$M for $k_B T \ll \Delta_0$.
We attribute these deviations to inter-ladder interactions and scattering of magnons from a dilute density of impurities.

The sample used for the experiments was a deuterated single crystal of IPA-CuCl$_3$ of volume $\sim 3$~g prepared by a solution growth method described elsewhere \cite{Manaka1997c}.
IPA-CuCl$_3$ crystallizes in the triclinic space group $P-1$ \cite{Manaka1997c}.
We index the momentum-space coordinates $\mathbf{q} = (h,k,l)$ in the corresponding reciprocal lattice units (r.l.u.).
The nonmagnetic IPA molecules form layers in the \textit{a-c} plane so that the magnetic interactions along \textit{b} are negligible \cite{Masuda2006}.
The Cu$^{2+}$ ions carry a spin $S=1/2$ and form ladders along the \textit{a} axis (Fig.~\ref{fig:IPA}~a).
As discussed in detail elsewhere \cite{Masuda2006}, exchange interactions within these ladders generate a quantum spin gap of magnitude 1.17~meV at $\mathbf{q} = (0.5,0,0)$.
The magnon dispersion \cite{Masuda2006} is represented as a color plot in Fig.~\ref{fig:IPA}~b.
The primary dispersion along \textit{h} is well reproduced by a spin-ladder model with ferromagnetic coupling $J_1 = -2.3$~meV along the rungs and antiferromagnetic couplings $J_2 = 1.2$~meV and $ J_3=2.9 $~meV along the legs and diagonals of the ladders respectively.
Inter-ladder interactions of magnitude $J_4 \sim -0.3$~meV generate a weak dispersion along the \textit{l}-direction \cite{Fischer2010}.

The NRSE-TAS experiments were performed on the spectrometer TRISP at the research reactor FRM-II in Garching, Germany \cite{Keller2002}.
The incident neutron beam was spin-polarized by a supermirror guide.
The scattered beam polarization with fixed $k_f=2.15$~$\AA^{-1}$ was measured with a transmission polarizer in front of the detector.
A velocity selector was used to cut out higher-order contamination of the incident beam.
The (002) reflection of pyrolytic-graphite was used to select the energies of incident and scattered neutrons.

\begin{figure}
	\includegraphics[width=8.5cm]{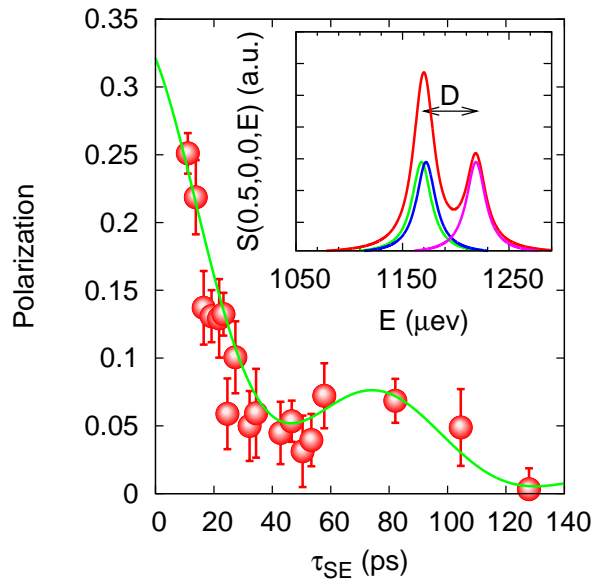}
    \caption{(Color online) Neutron beam polarization as a function of spin echo time measured at $T=0.5$~K at the bottom of the magnon dispersion at $\mathbf{q} = (0.5,0,0)$. The line is the result of a fit to three Lorentzian modes. The inset shows the spectrum extracted from the fit. \label{fig:0p5K:LW}}
\end{figure}

Figure~\ref{fig:0p5K:LW} shows typical NRSE data at the bottom of the magnon dispersion relation at momentum $\mathbf{q} = (0.5,0,0)$ and temperature $T=0.5$~K.
The dependence of the polarization $P$ of the detected neutron beam on the spin-echo time $\tau_{SE}$ is proportional to the Fourier transform of the magnon line shape \cite{Keller2003}.
As demonstrated previously for antiferromagnetic magnons \cite{Bayrakci2006} and phonons in metals \cite{Keller2006,Aynajian2008}, a Lorentzian lineshape thus yields an exponential $P (\tau_{SE})$ profile.
Clearly, the non-monotonic profile shown on Fig.~\ref{fig:0p5K:LW} is inconsistent with a single Lorentzian mode.
Rather, the profile is well described by a superposition of two modes with an energy difference $D = 37 $~$ \mu$eV and an intensity ratio $\sim 1:2$ (inset in Fig.~\ref{fig:0p5K:LW}).
This splitting could not be resolved in prior INS experiments with TAS \cite{Masuda2006,Zheludev2007,Zheludev2008}, but our data are consistent with ESR experiments at higher temperatures that indicated a splitting of the $S=1$ magnon resonance into $S^z=0$ and $S^z= \pm 1$ components \cite{Manaka2000}.
This small splitting is probably due to anisotropic superexchange interactions between Cu$^{2+}$ spins that act as a single-ion anisotropy on the effective $ S=1 $ spin on the rung of the ladder.
A detailed analysis shows that the lower mode arising from $S^z= \pm 1$ excitations is broader than the upper-energy $S^z=0$ mode at all temperatures.
This broadening may be the result of an additional splitting by $\sim 5$~$\mu$eV that is not resolved because it is comparable to the intrinsic widths of the individual lines (see below).
A small additional splitting due to dipolar and/or anisotropic exchange interactions is consistent with the low symmetry of the crystal lattice and with prior ESR data that showed a shift of the so-called ``half-field resonance'' away from the magnetic field expected without this splitting \cite{Manaka2000}.
However, since we were unable to definitively resolve the origin of the effective broadening of the lower-energy mode, we focus on the non-degenerate higher-energy mode from now on.

\begin{figure}
	\includegraphics[width=8.5cm]{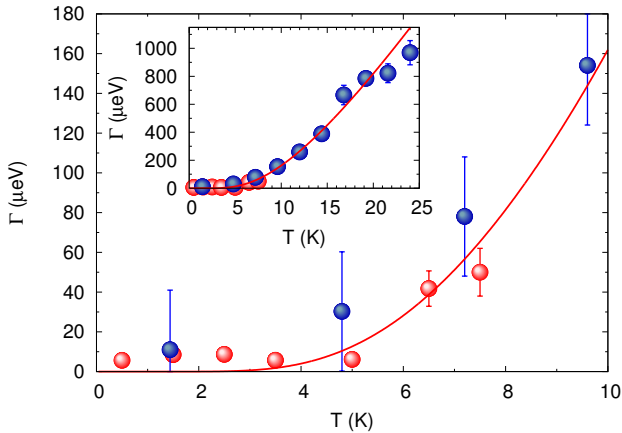}
    \caption{(Color online) Temperature dependence of the magnon line width $\Gamma$. Red (light) symbols are results of the current NRSE-TAS study. Blue (dark) symbols are TAS results from Ref.~\onlinecite{Zheludev2008} assuming one Lorentzian mode only. The line represents NL$\sigma$M calculations \cite{Damle1998} with $\Delta_0=1.44$~meV. Below $ T=5 $~K, the magnon modes are significantly broadened relative to the theoretical predictions. Error-bars on the data below 6~K are smaller than 2~$\mu$eV. The inset shows the temperature dependence of $\Gamma$ over a broad temperature range.  \label{fig:LW}}
\end{figure}

Figure~\ref{fig:LW} shows the linewidth $\Gamma$ of this mode extracted from fits to NRSE profiles over a wide temperature range.
At temperatures $T \gtrsim 5$ K, where $\Gamma$ exceeds the splitting $D$ between the modes, our data are in good agreement with the analysis of prior INS data that assumed a single Lorentzian profile.
As noted earlier, the NL$\sigma$M yields an adequate description of $\Gamma(T)$ in this temperature range.
At lower temperatures, however, the NRSE data show a saturation of $\Gamma$ at a value of $\sim 10$~$\mu$eV, in clear disagreement with the NL$\sigma$M that predicts an exponential reduction of $\Gamma$ as $T \rightarrow 0$ (line in Fig.~\ref{fig:LW}).
The temperature independence of $\Gamma$ at low temperatures indicates that the magnon lifetime is limited by scattering from impurities.
By assuming that $\Gamma$ becomes $T$-independent when the mean distance between thermally excited magnons equals the mean distance between impurities, we obtain a magnon mean free path of $\sim 200$ lattice spacings at $T=0.5$ K, and hence a rough estimate of 0.5\% for the impurity concentration.
Since this level of impurity contamination is difficult to avoid in real materials, these results indicate a generic limitation of experimental tests of the NL$\sigma$M predictions.
Our spectroscopic results agree with the conclusions of recent heat transport studies of nominally pure spin-chain and spin-ladder compounds \cite{Kordonis2006,Hess2007,Sologubenko2008}.
We cannot rule out, however, that the subtle broadening reflects residual interactions among magnons imposed by long-range, dipolar interactions \cite{Syromyatnikov2010}.
Further theoretical work is required to discriminate between these scenarios.

\begin{figure}
	\includegraphics[angle=-90,width=8.5cm]{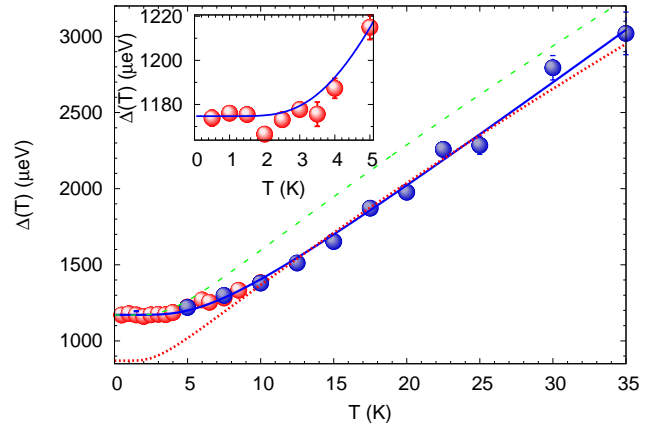}
    \caption{(Color online) Magnon gap energy as a function of temperature. Red (light) symbols are the current NRSE-TAS results. Blue (dark) symbols mark data points from Ref.~\onlinecite{Zheludev2008}. The red dotted and green dashed lines are result of the 1D-NL$\sigma$M with $\Delta_0 = 0.87$~meV and $\Delta_0 = 1.17$~meV respectively. The blue line is the result of a calculation based on a NL$\sigma$M with $\Delta_0 = 1.44$~meV and inter-ladder interaction $J_\perp / J =0.1$, following Ref.~\onlinecite{Senechal1993} as described in the text. The inset shows a blowup of the NRSE-TAS data set at low temperatures. Note the small ($\sim 8$~$\mu$eV) dip at $T=2$~K. The error-bars are smaller than the symbols. \label{fig:GapT}}
\end{figure}

The NRSE-TAS profiles also yield highly accurate measurements of the temperature dependent shift of the spin gap $\Delta$ relative to the zero-temperature value $\Delta_0 = 1.17$~meV determined by TAS.
Figure~\ref{fig:GapT} shows $\Delta(T)$ extracted from fits to the NRSE data, along with prior TAS results \cite{Zheludev2008}.
The red doted and green dashed lines in the figure show that the pure, 1D-NL$\sigma$M fails to describe the $\Delta(T)$ dependence over the entire temperature range.
A possible origin of this disagreement is the small density of impurities we have inferred from the analysis of the line width.
Indeed, prior work on IPA-Cu(Cl$_{0.95}$Br$_{0.05}$)$_3$ has shown that defects can enhance $ \Delta_0$ relative to the pure system \cite{Hong2010}.
However, because of the minute impurity concentration and the small energy scale of the impurity-induced broadening of the resonance in our nominally pure sample, this effect can be ruled out as the origin of the deviations from the NL$\sigma$M predictions.
Likewise, the splitting of the resonance, which signals deviations from the isotropic Heisenberg interactions assumed by the NL$\sigma$M, is too small to be responsible for the disagreement with the NL$\sigma$M predictions.
We have explicitly verified that the temperature-induced shift of the higher-energy mode is nearly parallel to the one of the lower-energy mode shown in Fig.~\ref{fig:GapT}.
The splitting between both modes increases only slightly with increasing temperature (from $D = 37$~$\mu$eV at $T=0.5$~K to 55~$\mu$eV at 8~K).

Another possible origin of the discrepancy between our $\Delta(T)$ data and the 1D-NL$\sigma$M predictions is the inter-ladder exchange interaction $J_4$ that generates the $l$-dependent modulation of $\Delta$ shown in Fig.~\ref{fig:IPA}~b.
At temperatures $ J_4 \lesssim k_B T$, thermally excited magnons are expected to be confined to the 2D troughs of this dispersion, whereas for higher temperatures they are subject to the $l$-averaged, effectively 1D dispersion.
In order to test the viability of this dimensional-crossover scenario, we have fitted the full $\Delta(T)$ data set including prior TAS data to an expression resulting from a mean-field analysis of a spin-1 chain with intra-chain interaction $J$ on a cubic lattice \cite{Senechal1993}.
As shown in Fig.~\ref{fig:GapT}, this expression yields an excellent description of the entire data set for $\Delta_0 = 1.44$~meV and $J_\perp /J = 0.1$.
The value of $\Delta_0$ is different from the energy of the lowest-energy magnons for $T \rightarrow 0$ shown in Fig.~\ref{fig:GapT}, but it agrees very well with the $l$-averaged gap determined by INS which determines the magnon properties at high temperatures.
The same value results from the fit of the $\Gamma(T)$ data to the NL$\sigma$M shown in Fig.~\ref{fig:LW}.
The $J_\perp /J$ value extracted from the fit is not expected to be quantitatively accurate, because it is based on a quasi-3D model rather than the quasi-2D Hamiltonian describing the spin system in IPA-CuCl$_3$.
Nonetheless, it agrees within a factor of two with the exchange parameters extracted from the magnon dispersion \cite{Masuda2006}.
An explanation of the subtle dip of $\Delta(T)$ at $T \sim 2$~K suggested by the data in the inset of Fig.~\ref{fig:GapT} may require theoretical calculations based on a spin Hamiltonian specific to IPA-CuCl$_3$.

In summary, we have determined the intrinsic fine structure, linewidth, and energy of the magnon resonance in a model 1D quantum spin liquid over a wide temperature range.
As noted before \cite{Zheludev2008}, calculations based on a NL$\sigma$M for a pure system with isotropic 1D exchange interactions yield a surprisingly good description of the data at high temperatures.
At temperatures $T \ll \Delta_0 /k_B$, however, where magnons are expected to be good quasiparticles, we have found that spin-space and real-space anisotropies in the spin Hamiltonian as well as scattering of magnons from a dilute density of impurities induce substantial deviations from the predictions of this model.
These effects are generic to all experimental realizations of 1D model Hamiltonians and should therefore be taken into account in order to obtain quantitative descriptions of thermodynamic, thermal transport, and NMR data. We have shown that the spectroscopic information derived from NRSE-TAS experiments can provide a solid basis for such a fully realistic theory of quasi-1D quantum magnets.

\begin{acknowledgments}
We thank K. Buchner for technical assistance and the German DFG for financial support under grant No. SFB/TRR 80.
N.B. acknowledges support from the Prospective Research program No. PBELP2-125427 of the Swiss NSF.
H.M. is supported by a Grant-in-Aid for Young Scientist (B) JSPS.
\end{acknowledgments}


\begin{thebibliography}{25}%
\makeatletter
\providecommand \@ifxundefined [1]{%
 \@ifx{#1\undefined}
}%
\providecommand \@ifnum [1]{%
 \ifnum #1\expandafter \@firstoftwo
 \else \expandafter \@secondoftwo
 \fi
}%
\providecommand \@ifx [1]{%
 \ifx #1\expandafter \@firstoftwo
 \else \expandafter \@secondoftwo
 \fi
}%
\providecommand \natexlab [1]{#1}%
\providecommand \enquote  [1]{``#1''}%
\providecommand \bibnamefont  [1]{#1}%
\providecommand \bibfnamefont [1]{#1}%
\providecommand \citenamefont [1]{#1}%
\providecommand \href@noop [0]{\@secondoftwo}%
\providecommand \href [0]{\begingroup \@sanitize@url \@href}%
\providecommand \@href[1]{\@@startlink{#1}\@@href}%
\providecommand \@@href[1]{\endgroup#1\@@endlink}%
\providecommand \@sanitize@url [0]{\catcode `\\12\catcode `\$12\catcode
  `\&12\catcode `\#12\catcode `\^12\catcode `\_12\catcode `\%12\relax}%
\providecommand \@@startlink[1]{}%
\providecommand \@@endlink[0]{}%
\providecommand \url  [0]{\begingroup\@sanitize@url \@url }%
\providecommand \@url [1]{\endgroup\@href {#1}{\urlprefix }}%
\providecommand \urlprefix  [0]{URL }%
\providecommand \Eprint [0]{\href }%
\providecommand \doibase [0]{http://dx.doi.org/}%
\providecommand \selectlanguage [0]{\@gobble}%
\providecommand \bibinfo  [0]{\@secondoftwo}%
\providecommand \bibfield  [0]{\@secondoftwo}%
\providecommand \translation [1]{[#1]}%
\providecommand \BibitemOpen [0]{}%
\providecommand \bibitemStop [0]{}%
\providecommand \bibitemNoStop [0]{.\EOS\space}%
\providecommand \EOS [0]{\spacefactor3000\relax}%
\providecommand \BibitemShut  [1]{\csname bibitem#1\endcsname}%
\let\auto@bib@innerbib\@empty
\bibitem [{\citenamefont {Affleck}(1989)}]{Affleck1989}%
  \BibitemOpen
  \bibfield  {author} {\bibinfo {author} {\bibfnamefont {I.}~\bibnamefont
  {Affleck}},\ }\href@noop {} {\bibfield  {journal} {\bibinfo  {journal} {J.
  Phys.: Condens. Matter}\ }\textbf {\bibinfo {volume} {1}},\ \bibinfo {pages}
  {3047} (\bibinfo {year} {1989})}\BibitemShut {NoStop}%
\bibitem [{\citenamefont {Jolicoeur}\ and\ \citenamefont
  {Golinelli}(1994)}]{Jolicoeur1994}%
  \BibitemOpen
  \bibfield  {author} {\bibinfo {author} {\bibfnamefont {T.}~\bibnamefont
  {Jolicoeur}}\ and\ \bibinfo {author} {\bibfnamefont {O.}~\bibnamefont
  {Golinelli}},\ }\href@noop {} {\bibfield  {journal} {\bibinfo  {journal}
  {Phys. Rev. B}\ }\textbf {\bibinfo {volume} {50}},\ \bibinfo {pages} {9265}
  (\bibinfo {year} {1994})}\BibitemShut {NoStop}%
\bibitem [{\citenamefont {Sachdev}\ and\ \citenamefont
  {Damle}(1997)}]{Sachdev1997}%
  \BibitemOpen
  \bibfield  {author} {\bibinfo {author} {\bibfnamefont {S.}~\bibnamefont
  {Sachdev}}\ and\ \bibinfo {author} {\bibfnamefont {K.}~\bibnamefont
  {Damle}},\ }\href@noop {} {\bibfield  {journal} {\bibinfo  {journal} {Phys.
  Rev. Lett.}\ }\textbf {\bibinfo {volume} {78}},\ \bibinfo {pages} {943}
  (\bibinfo {year} {1997})}\BibitemShut {NoStop}%
\bibitem [{\citenamefont {Damle}\ and\ \citenamefont
  {Sachdev}(1998)}]{Damle1998}%
  \BibitemOpen
  \bibfield  {author} {\bibinfo {author} {\bibfnamefont {K.}~\bibnamefont
  {Damle}}\ and\ \bibinfo {author} {\bibfnamefont {S.}~\bibnamefont
  {Sachdev}},\ }\href@noop {} {\bibfield  {journal} {\bibinfo  {journal} {Phys.
  Rev. B}\ }\textbf {\bibinfo {volume} {57}},\ \bibinfo {pages} {8307}
  (\bibinfo {year} {1998})}\BibitemShut {NoStop}%
\bibitem [{\citenamefont {Kordonis}\ \emph {et~al.}(2006)\citenamefont
  {Kordonis}, \citenamefont {Sologubenko}, \citenamefont {Lorenz},
  \citenamefont {Cheong},\ and\ \citenamefont {Freimuth}}]{Kordonis2006}%
  \BibitemOpen
  \bibfield  {author} {\bibinfo {author} {\bibfnamefont {K.}~\bibnamefont
  {Kordonis}} \emph {et~al.}, }\href
  {\doibase 10.1103/PhysRevLett.97.115901} {\bibfield  {journal} {\bibinfo
  {journal} {Phys. Rev. Lett.}\ }\textbf {\bibinfo {volume} {97}},\ \bibinfo
  {pages} {115901} (\bibinfo {year} {2006}).}\BibitemShut {Stop}%
\bibitem [{\citenamefont {Hess}\ \emph {et~al.}(2007)\citenamefont {Hess},
  \citenamefont {ElHaes}, \citenamefont {Waske}, \citenamefont {B\"uchner},
  \citenamefont {Sekar}, \citenamefont {Krabbes}, \citenamefont
  {Heidrich-Meisner},\ and\ \citenamefont {Brenig}}]{Hess2007}%
  \BibitemOpen
  \bibfield  {author} {\bibinfo {author} {\bibfnamefont {C.}~\bibnamefont
  {Hess}} \emph {et~al.},\ }\href {\doibase
  10.1103/PhysRevLett.98.027201} {\bibfield  {journal} {\bibinfo  {journal}
  {Phys. Rev. Lett.}\ }\textbf {\bibinfo {volume} {98}},\ \bibinfo {pages}
  {027201} (\bibinfo {year} {2007})}\BibitemShut {NoStop}%
\bibitem [{\citenamefont {Sologubenko}\ \emph {et~al.}(2008)\citenamefont
  {Sologubenko}, \citenamefont {Lorenz}, \citenamefont {Mydosh}, \citenamefont
  {Rosch}, \citenamefont {Shortsleeves},\ and\ \citenamefont
  {Turnbull}}]{Sologubenko2008}%
  \BibitemOpen
  \bibfield  {author} {\bibinfo {author} {\bibfnamefont {A.~V.}\ \bibnamefont
  {Sologubenko}} \emph {et~al.},\ }\href {\doibase
  10.1103/PhysRevLett.100.137202} {\bibfield  {journal} {\bibinfo  {journal}
  {Phys. Rev. Lett.}\ }\textbf {\bibinfo {volume} {100}},\ \bibinfo {pages}
  {137202} (\bibinfo {year} {2008})}\BibitemShut {NoStop}%
\bibitem [{\citenamefont {Takigawa}\ \emph {et~al.}(1996)\citenamefont
  {Takigawa}, \citenamefont {Asano}, \citenamefont {Ajiro}, \citenamefont
  {Mekata},\ and\ \citenamefont {Uemura}}]{Takigawa1996}%
  \BibitemOpen
  \bibfield  {author} {\bibinfo {author} {\bibfnamefont {M.}~\bibnamefont
  {Takigawa}}, \bibinfo {author} {\bibfnamefont {T.}~\bibnamefont {Asano}},
  \bibinfo {author} {\bibfnamefont {Y.}~\bibnamefont {Ajiro}}, \bibinfo
  {author} {\bibfnamefont {M.}~\bibnamefont {Mekata}}, \ and\ \bibinfo {author}
  {\bibfnamefont {Y.~J.}\ \bibnamefont {Uemura}},\ }\href {\doibase
  10.1103/PhysRevLett.76.2173} {\bibfield  {journal} {\bibinfo  {journal}
  {Phys. Rev. Lett.}\ }\textbf {\bibinfo {volume} {76}},\ \bibinfo {pages}
  {2173} (\bibinfo {year} {1996})}\BibitemShut {NoStop}%
\bibitem [{\citenamefont {Ghoshray}\ \emph {et~al.}(2005)\citenamefont
  {Ghoshray}, \citenamefont {Pahari}, \citenamefont {Bandyopadhyay},
  \citenamefont {Sarkar},\ and\ \citenamefont {Ghoshray}}]{Ghoshray2005}%
  \BibitemOpen
  \bibfield  {author} {\bibinfo {author} {\bibfnamefont {K.}~\bibnamefont
  {Ghoshray}}, \bibinfo {author} {\bibfnamefont {B.}~\bibnamefont {Pahari}},
  \bibinfo {author} {\bibfnamefont {B.}~\bibnamefont {Bandyopadhyay}}, \bibinfo
  {author} {\bibfnamefont {R.}~\bibnamefont {Sarkar}}, \ and\ \bibinfo {author}
  {\bibfnamefont {A.}~\bibnamefont {Ghoshray}},\ }\href {\doibase
  10.1103/PhysRevB.71.214401} {\bibfield  {journal} {\bibinfo  {journal} {Phys.
  Rev. B}\ }\textbf {\bibinfo {volume} {71}},\ \bibinfo {pages} {214401}
  (\bibinfo {year} {2005})}\BibitemShut {NoStop}%
\bibitem [{\citenamefont {Manaka}\ and\ \citenamefont
  {Yamada}(2000)}]{Manaka2000}%
  \BibitemOpen
  \bibfield  {author} {\bibinfo {author} {\bibfnamefont {H.}~\bibnamefont
  {Manaka}}\ and\ \bibinfo {author} {\bibfnamefont {I.}~\bibnamefont
  {Yamada}},\ }\href@noop {} {\bibfield  {journal} {\bibinfo  {journal} {Phys.
  Rev. B}\ }\textbf {\bibinfo {volume} {62}},\ \bibinfo {pages} {14279}
  (\bibinfo {year} {2000})}\BibitemShut {NoStop}%
\bibitem [{\citenamefont {Xu}\ \emph {et~al.}(2007)\citenamefont {Xu},
  \citenamefont {Broholm}, \citenamefont {Soh}, \citenamefont {Aeppli},
  \citenamefont {DiTusa}, \citenamefont {Chen}, \citenamefont {Kenzelmann},
  \citenamefont {Frost}, \citenamefont {Ito}, \citenamefont {Oka},\ and\
  \citenamefont {Takagi}}]{Xu2007}%
  \BibitemOpen
  \bibfield  {author} {\bibinfo {author} {\bibfnamefont {G.~Y.}\ \bibnamefont
  {Xu}} \emph {et~al.},\ }\href@noop {}
  {\bibfield  {journal} {\bibinfo  {journal} {Science}\ }\textbf {\bibinfo
  {volume} {317}},\ \bibinfo {pages} {1049} (\bibinfo {year}
  {2007})}\BibitemShut {NoStop}%
\bibitem [{\citenamefont {Zheludev}\ \emph {et~al.}(2008)\citenamefont
  {Zheludev}, \citenamefont {Garlea}, \citenamefont {Regnault}, \citenamefont
  {Manaka}, \citenamefont {Tsvelik},\ and\ \citenamefont
  {Chung}}]{Zheludev2008}%
  \BibitemOpen
  \bibfield  {author} {\bibinfo {author} {\bibfnamefont {A.}~\bibnamefont
  {Zheludev}} \emph {et~al.},\ }\href@noop
  {} {\bibfield  {journal} {\bibinfo  {journal} {Phys. Rev. Lett.}\ }\textbf
  {\bibinfo {volume} {100}},\ \bibinfo {pages} {157204} (\bibinfo {year}
  {2008})}\BibitemShut {NoStop}%
\bibitem [{\citenamefont {Bayrakci}\ \emph {et~al.}(2006)\citenamefont
  {Bayrakci}, \citenamefont {Keller}, \citenamefont {Habicht},\ and\
  \citenamefont {Keimer}}]{Bayrakci2006}%
  \BibitemOpen
  \bibfield  {author} {\bibinfo {author} {\bibfnamefont {S.~P.}\ \bibnamefont
  {Bayrakci}}, \bibinfo {author} {\bibfnamefont {T.}~\bibnamefont {Keller}},
  \bibinfo {author} {\bibfnamefont {K.}~\bibnamefont {Habicht}}, \ and\
  \bibinfo {author} {\bibfnamefont {B.}~\bibnamefont {Keimer}},\ }\href@noop {}
  {\bibfield  {journal} {\bibinfo  {journal} {Science}\ }\textbf {\bibinfo
  {volume} {312}},\ \bibinfo {pages} {1926} (\bibinfo {year}
  {2006})}\BibitemShut {NoStop}%
\bibitem [{\citenamefont {Keller}\ \emph {et~al.}(2006)\citenamefont {Keller},
  \citenamefont {Aynajian}, \citenamefont {Habicht}, \citenamefont {Boeri},
  \citenamefont {Bose},\ and\ \citenamefont {Keimer}}]{Keller2006}%
  \BibitemOpen
  \bibfield  {author} {\bibinfo {author} {\bibfnamefont {T.}~\bibnamefont
  {Keller}}, \bibinfo {author} {\bibfnamefont {P.}~\bibnamefont {Aynajian}},
  \bibinfo {author} {\bibfnamefont {K.}~\bibnamefont {Habicht}}, \bibinfo
  {author} {\bibfnamefont {L.}~\bibnamefont {Boeri}}, \bibinfo {author}
  {\bibfnamefont {S.~K.}\ \bibnamefont {Bose}}, \ and\ \bibinfo {author}
  {\bibfnamefont {B.}~\bibnamefont {Keimer}},\ }\href {\doibase
  10.1103/PhysRevLett.96.225501} {\bibfield  {journal} {\bibinfo  {journal}
  {Phys. Rev. Lett.}\ }\textbf {\bibinfo {volume} {96}},\ \bibinfo {pages}
  {225501} (\bibinfo {year} {2006})}\BibitemShut {NoStop}%
\bibitem [{\citenamefont {Aynajian}\ \emph {et~al.}(2008)\citenamefont
  {Aynajian}, \citenamefont {Keller}, \citenamefont {Boeri}, \citenamefont
  {Shapiro}, \citenamefont {Habicht},\ and\ \citenamefont
  {Keimer}}]{Aynajian2008}%
  \BibitemOpen
  \bibfield  {author} {\bibinfo {author} {\bibfnamefont {P.}~\bibnamefont
  {Aynajian}}, \bibinfo {author} {\bibfnamefont {T.}~\bibnamefont {Keller}},
  \bibinfo {author} {\bibfnamefont {L.}~\bibnamefont {Boeri}}, \bibinfo
  {author} {\bibfnamefont {S.~M.}\ \bibnamefont {Shapiro}}, \bibinfo {author}
  {\bibfnamefont {K.}~\bibnamefont {Habicht}}, \ and\ \bibinfo {author}
  {\bibfnamefont {B.}~\bibnamefont {Keimer}},\ }\href {\doibase
  10.1126/science.1154115} {\bibfield  {journal} {\bibinfo  {journal}
  {Science}\ }\textbf {\bibinfo {volume} {319}},\ \bibinfo {pages} {1509}
  (\bibinfo {year} {2008})}\BibitemShut {NoStop}%
\bibitem [{\citenamefont {Masuda}\ \emph {et~al.}(2006)\citenamefont {Masuda},
  \citenamefont {Zheludev}, \citenamefont {Manaka}, \citenamefont {Regnault},
  \citenamefont {Chung},\ and\ \citenamefont {Qiu}}]{Masuda2006}%
  \BibitemOpen
  \bibfield  {author} {\bibinfo {author} {\bibfnamefont {T.}~\bibnamefont
  {Masuda}} \emph {et~al.},\ }\href@noop {} {\bibfield
  {journal} {\bibinfo  {journal} {Phys. Rev. Lett.}\ }\textbf {\bibinfo
  {volume} {96}},\ \bibinfo {pages} {047210} (\bibinfo {year}
  {2006})}\BibitemShut {NoStop}%
\bibitem [{\citenamefont {Zheludev}\ \emph {et~al.}(2007)\citenamefont
  {Zheludev}, \citenamefont {Garlea}, \citenamefont {Masuda}, \citenamefont
  {Manaka}, \citenamefont {Regnault}, \citenamefont {Ressouche}, \citenamefont
  {Grenier}, \citenamefont {Chung}, \citenamefont {Qiu}, \citenamefont
  {Habicht}, \citenamefont {Kiefer},\ and\ \citenamefont
  {Boehm}}]{Zheludev2007}%
  \BibitemOpen
  \bibfield  {author} {\bibinfo {author} {\bibfnamefont {A.}~\bibnamefont
  {Zheludev}} \emph {et~al.},\ }\href
  {\doibase 10.1103/PhysRevB.76.054450} {\bibfield  {journal} {\bibinfo
  {journal} {Phys. Rev. B}\ }\textbf {\bibinfo {volume} {76}},\ \bibinfo
  {pages} {054450} (\bibinfo {year} {2007})}\BibitemShut {NoStop}%
\bibitem [{\citenamefont {Garlea}\ \emph {et~al.}(2007)\citenamefont {Garlea},
  \citenamefont {Zheludev}, \citenamefont {Masuda}, \citenamefont {Manaka},
  \citenamefont {Regnault}, \citenamefont {Ressouche}, \citenamefont {Grenier},
  \citenamefont {Chung}, \citenamefont {Qiu}, \citenamefont {Habicht},
  \citenamefont {Kiefer},\ and\ \citenamefont {Boehm}}]{Garlea2007}%
  \BibitemOpen
  \bibfield  {author} {\bibinfo {author} {\bibfnamefont {V.~O.}\ \bibnamefont
  {Garlea}} \emph {et~al.},\ }\href
  {\doibase 10.1103/PhysRevLett.98.167202} {\bibfield  {journal} {\bibinfo
  {journal} {Phys. Rev. Lett.}\ }\textbf {\bibinfo {volume} {98}},\ \bibinfo
  {pages} {167202} (\bibinfo {year} {2007})}\BibitemShut {NoStop}%
\bibitem [{\citenamefont {Manaka}\ \emph {et~al.}(1997)\citenamefont {Manaka},
  \citenamefont {Yamada},\ and\ \citenamefont {Yamaguchi}}]{Manaka1997c}%
  \BibitemOpen
  \bibfield  {author} {\bibinfo {author} {\bibfnamefont {H.}~\bibnamefont
  {Manaka}}, \bibinfo {author} {\bibfnamefont {I.}~\bibnamefont {Yamada}}, \
  and\ \bibinfo {author} {\bibfnamefont {K.}~\bibnamefont {Yamaguchi}},\
  }\href@noop {} {\bibfield  {journal} {\bibinfo  {journal} {J. Phys. Soc.
  Jpn.}\ }\textbf {\bibinfo {volume} {66}},\ \bibinfo {pages} {564} (\bibinfo
  {year} {1997})}\BibitemShut {NoStop}%
\bibitem [{\citenamefont {Fischer}\ \emph {et~al.}(2010)\citenamefont
  {Fischer}, \citenamefont {Duffe},\ and\ \citenamefont {Uhrig}}]{Fischer2010}%
  \BibitemOpen
  \bibfield  {author} {\bibinfo {author} {\bibfnamefont {T.}~\bibnamefont
  {Fischer}}, \bibinfo {author} {\bibfnamefont {S.}~\bibnamefont {Duffe}}, \
  and\ \bibinfo {author} {\bibfnamefont {G.~S.}~\bibnamefont {Uhrig}},\
  }\href@noop {} {\bibfield  {journal} {\bibinfo  {journal}
  {arXiv:1009.3375v1}\ } (\bibinfo {year} {2010})}\BibitemShut {NoStop}%
\bibitem [{\citenamefont {Keller}\ \emph {et~al.}(2002)\citenamefont {Keller},
  \citenamefont {Habicht}, \citenamefont {Klann}, \citenamefont {Schneider},\
  and\ \citenamefont {Keimer}}]{Keller2002}%
  \BibitemOpen
  \bibfield  {author} {\bibinfo {author} {\bibfnamefont {T.}~\bibnamefont
  {Keller}}, \bibinfo {author} {\bibfnamefont {K.}~\bibnamefont {Habicht}},
  \bibinfo {author} {\bibfnamefont {H.}~\bibnamefont {Klann}}, \bibinfo
  {author} {\bibfnamefont {H.}~\bibnamefont {Schneider}}, \ and\ \bibinfo
  {author} {\bibfnamefont {B.}~\bibnamefont {Keimer}},\ }\href@noop {}
  {\bibfield  {journal} {\bibinfo  {journal} {Appl. Phys. A}\ }\textbf
  {\bibinfo {volume} {74}},\ \bibinfo {pages} {s332} (\bibinfo {year}
  {2002})}\BibitemShut {NoStop}%
\bibitem [{\citenamefont {Keller}\ \emph {et~al.}(2003)\citenamefont {Keller},
  \citenamefont {Keimer}, \citenamefont {Habicht}, \citenamefont {Golub},\ and\
  \citenamefont {Mezei}}]{Keller2003}%
  \BibitemOpen
  \bibfield  {author} {\bibinfo {author} {\bibfnamefont {T.}~\bibnamefont
  {Keller}}, \bibinfo {author} {\bibfnamefont {B.}~\bibnamefont {Keimer}},
  \bibinfo {author} {\bibfnamefont {K.}~\bibnamefont {Habicht}}, \bibinfo
  {author} {\bibfnamefont {R.}~\bibnamefont {Golub}}, \ and\ \bibinfo {author}
  {\bibfnamefont {F.}~\bibnamefont {Mezei}},\ }in\ \href@noop {} {\emph
  {\bibinfo {booktitle} {Neutron Spin Echo Spectroscopy}}},\ \bibinfo {series}
  {Lecture Notes in Physics}, Vol.\ \bibinfo {volume} {601},\ \bibinfo {editor}
  {edited by\ \bibinfo {editor} {\bibfnamefont {F.}~\bibnamefont {Mezei}},
  \bibinfo {editor} {\bibfnamefont {C.}~\bibnamefont {Pappas}}, \ and\ \bibinfo
  {editor} {\bibfnamefont {T.}~\bibnamefont {Gutberlet}}}\ (\bibinfo
  {publisher} {Springer Berlin / Heidelberg},\ \bibinfo {year} {2003})\ pp.\
  \bibinfo {pages} {74--86}\BibitemShut {NoStop}%
\bibitem [{\citenamefont {Syromyatnikov}(2010)}]{Syromyatnikov2010}%
  \BibitemOpen
  \bibfield  {author} {\bibinfo {author} {\bibfnamefont {A.~V.}\ \bibnamefont
  {Syromyatnikov}},\ }\href {\doibase 10.1103/PhysRevB.82.024432} {\bibfield
  {journal} {\bibinfo  {journal} {Phys. Rev. B}\ }\textbf {\bibinfo {volume}
  {82}},\ \bibinfo {pages} {024432} (\bibinfo {year} {2010})}\BibitemShut
  {NoStop}%
\bibitem [{\citenamefont {Senechal}(1993)}]{Senechal1993}%
  \BibitemOpen
  \bibfield  {author} {\bibinfo {author} {\bibfnamefont {D.}~\bibnamefont
  {Senechal}},\ }\href@noop {} {\bibfield  {journal} {\bibinfo  {journal}
  {Phys. Rev. B}\ }\textbf {\bibinfo {volume} {48}},\ \bibinfo {pages} {15880}
  (\bibinfo {year} {1993})}\BibitemShut {NoStop}%
\bibitem [{\citenamefont {Hong}\ \emph {et~al.}(2010)\citenamefont {Hong},
  \citenamefont {Zheludev}, \citenamefont {Manaka},\ and\ \citenamefont
  {Regnault}}]{Hong2010}%
  \BibitemOpen
  \bibfield  {author} {\bibinfo {author} {\bibfnamefont {T.}~\bibnamefont
  {Hong}}, \bibinfo {author} {\bibfnamefont {A.}~\bibnamefont {Zheludev}},
  \bibinfo {author} {\bibfnamefont {H.}~\bibnamefont {Manaka}}, \ and\ \bibinfo
  {author} {\bibfnamefont {L.-P.}\ \bibnamefont {Regnault}},\ }\href@noop {}
  {\bibfield  {journal} {\bibinfo  {journal} {Phys. Rev. B}\ }\textbf {\bibinfo
  {volume} {81}},\ \bibinfo {pages} {060410(R)} (\bibinfo {year}
  {2010})}\BibitemShut {NoStop}%
\end{thebibliography}

%

\end{document}